\documentclass[a4paper,12pt]{article}
\pdfoutput=1
\usepackage{graphicx,rotating,amssymb,amsmath,cite}
\usepackage[normalem]{ulem}

\ifx\pdfoutput\undefined
\usepackage[dvips,bookmarks]{hyperref}	
\else
\usepackage{hyperref}	
\fi
\hypersetup{colorlinks,bookmarksopen,bookmarksnumbered,citecolor=rossos,
linkcolor=verde,pdfstartview=FitH,urlcolor=rossos}

\def\hhref#1{\href{http://arxiv.org/abs/#1}{#1}} 

\usepackage{footnote}
\usepackage{booktabs}
\usepackage{multicol}
\usepackage{bold-extra}
\usepackage{color}
\definecolor{rosso}{cmyk}{0,1,1,0.4}
\definecolor{rossos}{cmyk}{0,1,1,0.55}
\definecolor{rossoc}{cmyk}{0,1,1,0.2}
\definecolor{blu}{cmyk}{1,1,0,0.3}
\definecolor{blus}{cmyk}{1,1,0,0.6}
\definecolor{bluc}{cmyk}{1,1,0,0.1}
\definecolor{verde}{cmyk}{0.92,0,0.59,0.25}
\definecolor{verdec}{cmyk}{0.92,0,0.59,0.15}
\definecolor{verdes}{cmyk}{0.92,0,0.59,0.4}

\font\tenrsfs=rsfs10 at 12pt
\font\sevenrsfs=rsfs7
\font\fiversfs=rsfs5
\newfam\rsfsfam
\textfont\rsfsfam=\tenrsfs
\scriptfont\rsfsfam=\sevenrsfs
\scriptscriptfont\rsfsfam=\fiversfs
\def\mathscr#1{{\fam\rsfsfam\relax#1}}

\def\circa#1{\,\raise.3ex\hbox{$#1$\kern-.75em\lower1ex\hbox{$\sim$}}\,}

\newcommand{\beq}{\begin{equation}}
\newcommand{\eeq}{\end{equation}}

\def\circa#1{\,\raise.3ex\hbox{$#1$\kern-.75em\lower1ex\hbox{$\sim$}}\,}
\makeatletter

\def\art{\@ifnextchar[{\eart}{\oart}}
\def\eart[#1]#2#3#4#5#6{{\rm #2}, {#3 #4} {\rm (#6) #5} [{\hhref{#1}}]}
\def\hepart[#1]#2{{\rm #2, \hhref{#1}}}
\newcommand{\oart}[5]{{\rm #1}, {#2 #3} {\rm (#5) #4}}

\newcounter{alphaequation}[equation]
\def\thealphaequation{\theequation\hbox to
0.6em{\hfil\alph{alphaequation}\hfil}}
\def\eqnsystem#1{
\def\@eqnnum{{\rm (\thealphaequation)}}
\def\@@eqncr{\let\@tempa\relax \ifcase\@eqcnt \def\@tempa{& & &} \or
  \def\@tempa{& &}\or \def\@tempa{&}\fi\@tempa
  \if@eqnsw\@eqnnum\refstepcounter{alphaequation}\fi
\global\@eqnswtrue\global\@eqcnt=0\cr}
\refstepcounter{equation} \let\@currentlabel\theequation \def\@tempb{#1}
\ifx\@tempb\empty\else\label{#1}\fi
\refstepcounter{alphaequation}
\let\@currentlabel\thealphaequation
\global\@eqnswtrue\global\@eqcnt=0 \tabskip\@centering\let\\=\@eqncr
$$\halign to \displaywidth\bgroup \@eqnsel\hskip\@centering
$\displaystyle\tabskip\z@{##}$&\global\@eqcnt\@ne
\hskip2\arraycolsep\hfil${##}$\hfil& \global\@eqcnt\tw@\hskip2\arraycolsep
$\displaystyle\tabskip\z@{##}$\hfil
\tabskip\@centering&\llap{##}\tabskip\z@\cr}
\def\endeqnsystem{\@@eqncr\egroup$$\global\@ignoretrue} \makeatother

\begin{document}
\begin{flushright}
\footnotesize
\end{flushright}
\color{black}

\begin{center}
{\LARGE\bf X-rays constraints on sub-GeV Dark Matter}

\bigskip
\color{black}\vspace{0.6cm}

{
{\large\bf Jordan Koechler}
}
\\[3mm]
{\it \href{http://www.lpthe.jussieu.fr/spip/index.php}{Laboratoire de Physique Th\'eorique et Hautes Energies (LPTHE)},\\ UMR 7589 CNRS \& Sorbonne University, 4 Place Jussieu, F-75252, Paris, France}\\[3mm]
\end{center}

\bigskip

\centerline{\large\bf Abstract}

\begin{quote}

We present updated constraints on `light' Dark Matter (DM) particles with masses between 1 MeV and 5 GeV. In this range, we can expect DM-produced $e^\pm$ pairs to up-scatter low-energy ambient photons in the Milky Way via the Inverse Compton process, and produce a flux of X-rays that can be probed by a range of space observatories. Using diffuse X-ray data from {\sc Xmm-Newton}, {\sc Integral}, {\sc NuStar} and {\sc Suzaku}, we compute the strongest constraints to date on annihilating DM for 200 MeV $< m_{\rm DM} <$ 5 GeV and decaying DM for 100 MeV $< m_{\rm DM} <$ 5 GeV.

\end{quote}

\bigskip

\tableofcontents

\newpage

\section{Introduction}
\label{sec:introduction}

\bigskip

Among the greatest mysteries of the Universe, the Dark Matter (DM) problem is arguably the most puzzling one. Although DM constitutes around 25\% of the energy budget of the Universe today, almost a century of efforts was not enough to pinpoint its nature. On a brighter note, a few very well-motivated paradigms at different DM mass scales have been theorized, a textbook example being the weakly interacting massive particles (WIMPs). Thermal production of these particles provides the correct relic density for DM, and they naturally arise in several extensions of the Standard Model. However, because of the null results from direct, indirect and collider searches, the attention of the community slowly started to deviate from this paradigm.

Here we focus on lighter-than-WIMPs DM candidates. There is actually a whole class of light DM models that are well-motivated~\cite{Feng:2008ya,Hochberg:2014dra,Petraki:2013wwa}. Even though this trend starts to change, experiments that are dedicated to DM searches are mostly designed to probe WIMPs. In direct detection experiments, the sensitivity start to decrease significantly for DM masses around and below 1 GeV, and the need to study electronic recoils and the Migdal effect is becoming more and more important to overcome this sensitivity loss. The missing transverse energy -- holy grail of DM signature in colliders -- is tricky to reconstruct at low energies as instrumental background becomes dominant.

Probing light DM candidates in indirect detection is also a challenge for two reasons: i) low-energy charged particles produced by DM annihilation or decay in the Milky Way are unable to cross the heliopause, as solar winds push them away from the solar system and our detectors (except for {\sc Voyager} 1 and 2, from which a study of light DM has already been done~\cite{Boudaud:2016mos}), ii) there exists a lack of sensitivity in previous and current $\gamma$-ray observatories in the 100 keV -- 100 MeV energy range, named the `MeV gap'.

\medskip

In this proceeding, we present a novel approach that allows to circumvent the MeV gap by studying secondary emissions of X-rays from DM annihilations or decays in the Milky Way~\cite{Cirelli:2020bpc,Cirelli:2023tnx}. The picture is quite simple: considering the kinematically open DM annihilation or decay channels that will at some point produce $e^\pm$, we can study secondary emissions that are coming from the Inverse-Compton Scattering (ICS) of ambient photons on these $e^\pm$, producing X-rays that can be probed by a variety of observatories. We therefore perform a systematic analysis on available X-ray data from the {\sc Integral}~\cite{Winkler:2003nn,Bouchet:2011fn}, {\sc NuStar}~\cite{Harrison:2013md,Krivonos:2020qvl,Perez:2016tcq,Roach:2019ctw}, {\sc Suzaku}~\cite{Mitsuda:2007,Yoshino:2009} and {\sc Xmm-Newton}~\cite{XMM:2001haf,Foster:2021ngm} observatories, and in the end compute strong constraints on annihilating and decaying light DM. 

The rest of the proceeding is organized as follows: in sec.~\ref{sec:xrays} we explain the computation of the flux of X-rays from light DM annihilations or decays, in sec.~\ref{sec:analysis} we detail the analysis scheme in order to compute the constraints for all the datasets that we considered, and finally in sec.~\ref{sec:results} we show the main results and discuss them.


\section{X-rays from Dark Matter annihilations and decays}
\label{sec:xrays}
\bigskip

In this study, we focus on light DM with a mass range between 1 MeV and 5 GeV. This allows us to only study the following few annihilation or decay channels
\begin{eqnarray}
{\rm DM \, (DM)} &\longrightarrow &e^+ e^-, \label{eq:ee}\\
{\rm DM \, (DM)} &\longrightarrow &\mu^+ \mu^-, \label{eq:mumu}\\
{\rm DM \, (DM)} &\longrightarrow &\pi^+ \pi^-, \label{eq:pipi}
\end{eqnarray}
without considering any exotic channels involving hadronic or other mesonic resonances that can come from specific DM models. We do not consider annihilation or decay in neutral pions, since they decay into $\gamma$-rays that can either be probed by {\sc Fermi} (and therefore thoroughly studied~\cite{Essig:2013goa}) or be in the MeV gap. The same reasoning applies for ${\rm DM \, (DM) \to \gamma\gamma}$.

Given a specific channel, we consider the flux of photons coming from two contributions. First, prompt emissions which are photons produced during final state radiations (FSR) and radiative decays of the muons or charged pions (Rad). Second, and as mentioned above, the ICS of DM-produced $e^\pm$ -- either produced directly through eq.~\ref{eq:ee}, or from the decays of muons or charged pions from eqs.~\ref{eq:mumu} and~\ref{eq:pipi} -- on ambient photons in the Milky Way. Assuming DM is its own antiparticle, the differential flux of photons from prompt emissions is written
\begin{equation}
\frac{d\Phi_{f,\gamma}}{dE_\gamma d\Omega}=\frac{1}{4\pi} \frac{dN_{f,\gamma}}{dE_\gamma} \times \left\{
\begin{array}{cl}
\displaystyle
\frac{\langle \sigma v\rangle}{2} \int_{\rm l.o.s.}ds\,\left(\frac{\rho_{\rm DM}(r(s,\theta))}{m_{\rm DM}}\right)^2 &{\rm (annihilation)}\\
\\
\displaystyle
\ \; \Gamma \ \; \int_{\rm l.o.s.}ds\,\left(\frac{\rho_{\rm DM}(r(s,\theta))}{m_{\rm DM}}\right) &{\rm (decay)}
\end{array}
\right. ,
\end{equation}
where $dN_{f,\gamma}/dE_\gamma$ is the energy spectrum of photons from $f = \{{\rm FSR},{\rm Rad}\}$~\cite{Cirelli:2020bpc}. The spherically symmetric galactic DM energy density profile $\rho_{\rm DM}(r)$ (squared in the annihilation case) is integrated over the variable $s$ that runs along the line of sight (l.o.s.) which forms an angle $\theta$ with the Sun-Galactic Center (GC) axis. The differential flux of photons depends also on the DM mass $m_{\rm DM}$ and the annihilation cross section $\langle\sigma v\rangle$ (or decay rate $\Gamma = 1/\tau$ where $\tau$ is the decay half-time).

Computing the X-ray flux from the ICS is not as straight-forward. In order to do that, we need three ingredients. The first one is the local number density of DM-produced $e^\pm$, computed by solving semi-analytically the diffusion-loss equation for $e^\pm$ that arise from DM annihilations or decays and propagate in the galactic medium. Here we consider that muons and charged pions always produce $e^\pm$ at the end of their decay path. The second ingredient is the local number density of ambient photons. There are three sources of such photons in the galactic medium: the cosmic microwave background (CMB), dust-rescattered infrared (IR) light and optical starlight (SL). For the local number density of CMB photons we use a blackbody spectrum and for the two other sources we use the intensity maps from {\sc Galprop} v54~\cite{Vladimirov:2010aq} (in turn based on observations from {\sc Cobe/Dirbe}). Finally, the third ingredient is the Klein-Nishina cross section taken in the Thomson limit since the ambient photons have negligible energy compared to the DM-produced $e^\pm$. We invite the interested reader to check the full expression of the X-ray flux from ICS written in the main paper~\cite{Cirelli:2023tnx}.

\medskip

In the next section, we describe the analysis scheme of our study in order to produce the results shown in sec.~\ref{sec:results}.


\section{Analysis scheme}
\label{sec:analysis}
\bigskip

\begin{figure}[h]
\begin{minipage}{0.49\linewidth}
\centerline{\includegraphics[width=\linewidth]{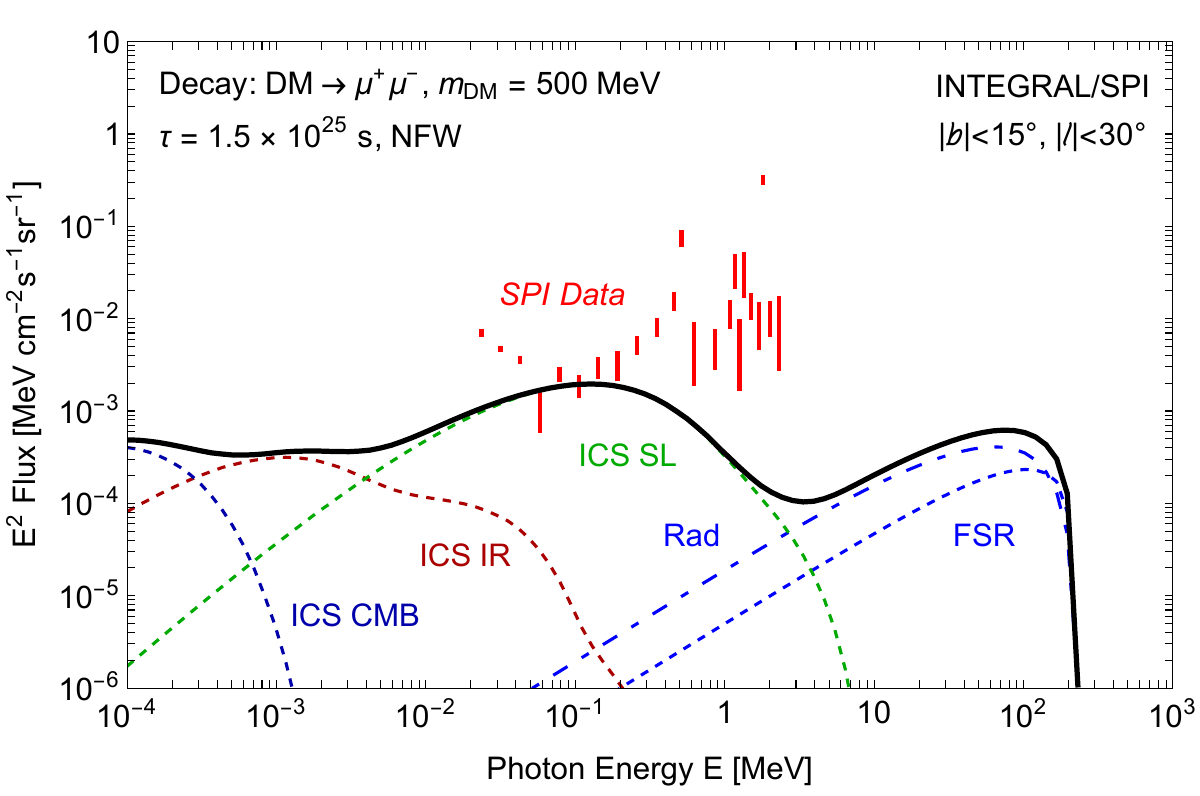}}
\end{minipage}
\hfill
\begin{minipage}{0.49\linewidth}
\centerline{\includegraphics[width=\linewidth]{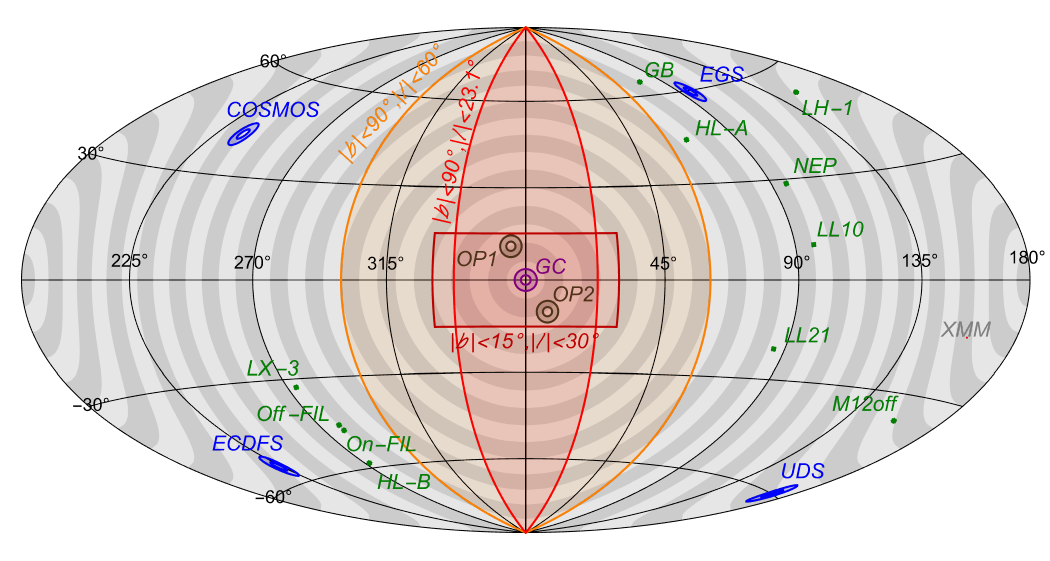}}
\end{minipage}
\caption[]{Left: Illustration of the X-ray fluxes from a specific DM decay channel and set of parameters, compared to measured flux from diffusion emission searches using {\sc Integral} in the given region of observation. Right: Illustrative map in galactic coordinates of the position of each field from which the data we use are taken from. The three regions in red and orange represent {\sc Integral} regions, the rings in blue, brown and purple {\sc NuStar} ones, the green fields represent {\sc Suzaku} regions and the rings in shades of gray {\sc Xmm-Newton}.}
\label{fig:illustration}
\end{figure}

Using the aforementioned ingredients, we can compute X-ray fluxes from DM annihilations or decays for a given region of observation in the Milky Way. For each channel, there are two free parameters: the DM annihilation cross section $\langle\sigma v\rangle$ (or decay rate $\Gamma$) and DM mass $m_{\rm DM}$. In order to set constraints on these parameters, we use a conservative approach: we do not attempt to predict any astrophysical background, and set the bound on $\langle\sigma v\rangle$ (or $\Gamma$) and $m_{\rm DM}$ when the predicted X-ray flux from DM exceeds the data we use. This can be translated in the test statistic
\begin{equation}
\chi_>^2 (p, m_{\rm DM})= \sum_{i \in {\rm bins}} \left(\frac{\max[\Phi_{{\rm DM}\gamma,i}(p, m_{\rm DM})-\phi_i,0]}{\sigma_i}\right)^2,
\label{eq:ts}
\end{equation}
where a 2$\sigma$ bound on $p=\langle\sigma v\rangle$ or $\Gamma$ is imposed for each value of $m_{\rm DM}$ whenever $\chi_>^2 \geq 4$, where $\Phi_{{\rm DM}\gamma,i}$ is the predicted X-ray flux, $\phi_i$ the measured flux and $\sigma_i$ its uncertainty in the data bin $i$. Left panel of fig.~\ref{fig:illustration} show an illustration of the X-ray fluxes we predict compared to one of the dataset we use in order to set the bounds.

We did a systematic analysis of available datasets from a variety of X-ray observatories: blank-sky fields~\cite{Krivonos:2020qvl}, GC~\cite{Perez:2016tcq} and off-plane~\cite{Roach:2019ctw} observations from {\sc NuStar}, diffuse emission searches from {\sc Integral}~\cite{Bouchet:2011fn}, high-latitude fields from {\sc Suzaku}~\cite{Yoshino:2009} and blank-sky data from {\sc Xmm-Newton}~\cite{Foster:2021ngm}. The right panel of fig.~\ref{fig:illustration} shows the regions of observation of each dataset we use. 

In the final section, we show the bounds we derive using eq.~\ref{eq:ts} for each aforementioned dataset, as well as comparing them to the current literature.


\section{Results}
\label{sec:results}
\bigskip

\begin{figure}[h]
\begin{minipage}{0.49\linewidth}
\centerline{\includegraphics[width=\linewidth]{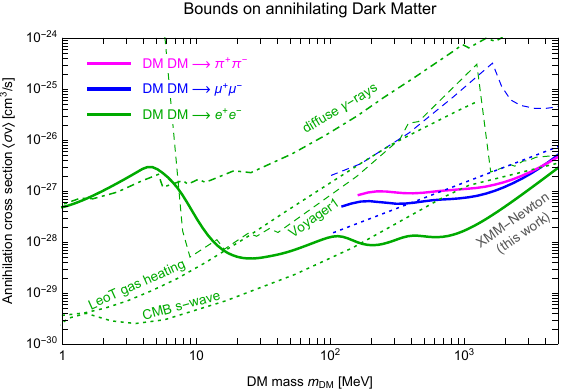}}
\end{minipage}
\hfill
\begin{minipage}{0.49\linewidth}
\centerline{\includegraphics[width=\linewidth]{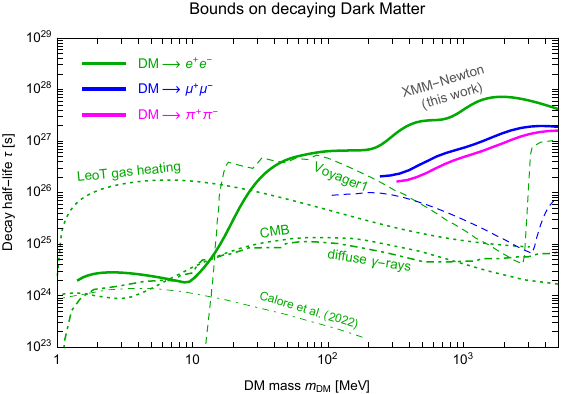}}
\end{minipage}
\hfill
\begin{minipage}{0.49\linewidth}
\centerline{\includegraphics[width=\linewidth]{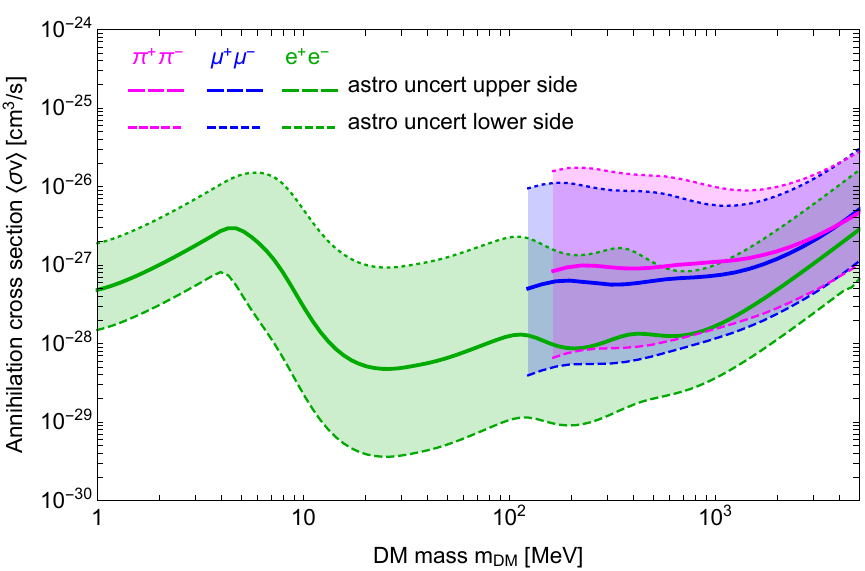}}
\end{minipage}
\hfill
\begin{minipage}{0.49\linewidth}
\centerline{\includegraphics[width=\linewidth]{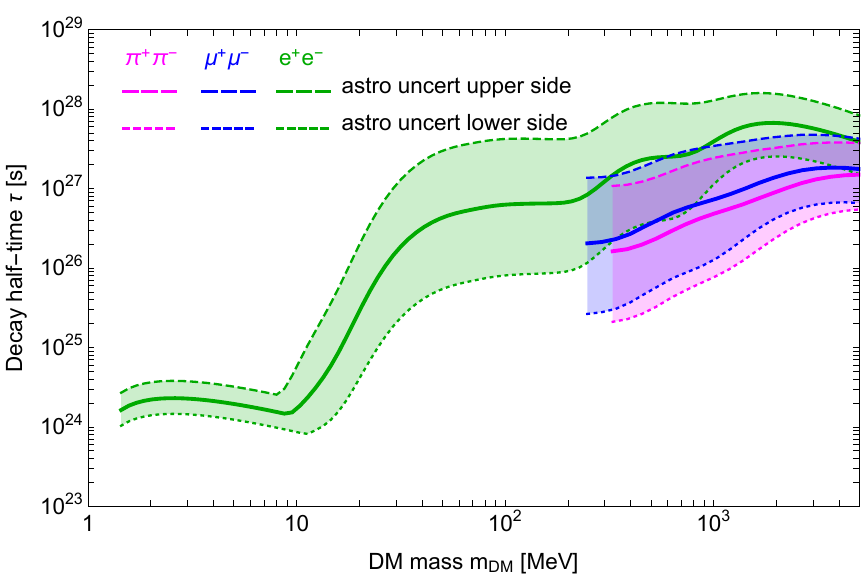}}
\end{minipage}
\caption[]{Top: Our main constraints on annihilating (left) and decaying (right) DM, in comparison with existing bounds. Bottom: Impact of the theoretical uncertainties on the bounds on annihilating (left) and decaying (right) DM we derived.}
\label{fig:mainbounds}
\end{figure}

In our study, we show that for every DM annihilation or decay channel and across the entire DM mass range, the bounds derived from the {\sc Xmm-Newton} dataset are the most stringent ones compared to other bounds derived using the remaining datasets. In the top panels of fig.~\ref{fig:mainbounds}, we plot our main bounds compared to other existing bounds in the same mass range: diffuse X- and $\gamma$-rays~\cite{Essig:2013goa} as a dot-dashed line, $e^\pm$ cosmic ray flux from outside the heliosphere using {\sc Voyager} 1 data~\cite{Boudaud:2016mos} as dashed lines, constraints on gas heating in the dwarf galaxy Leo T due to energy injection from s-wave DM annihilation or decay~\cite{Wadekar:2021qae}, impact on CMB anisotropies from the same energy injection~\cite{Slatyer:2015jla,Lopez-Honorez:2013cua,Liu:2016cnk}, the two latter ones as dotted lines, and finally FSR from decaying DM using 16 years of {\sc Integral} data~\cite{Calore:2022pks} as a thin dot-dashed line.

We have to stress that the diffuse X- and $\gamma$-rays bound did not include ICS emission. Our bounds are comparable to theirs in the DM mass range where ICS emission is sub-dominant, and are by orders of magnitude more stringent where the ICS emission dominates over the other X-ray emission processes. This indeed show that for higher DM masses, computing ICS emissions can be powerful for deriving strong bounds on light DM.

In the bottom panels of fig.~\ref{fig:mainbounds} we show the impact of theoretical uncertainties on our bounds. We considered four sources of uncertainty -- listed in decreasing order of importance: the choice of the galactic DM profile (our fiducial case uses NFW, but the profile could be cuspier or cored), the normalisation of the ambient photon intensity maps and of the galactic gas density that interact with the DM-produced $e^\pm$ (letting both vary by a factor of 2) and the choice of galactic magnetic field profile. The uncertainty bands can span up to two orders of magnitude.

Considering our fiducial case (as plotted in the top panels of fig.~\ref{fig:mainbounds}), the bounds we derived are the most stringent ones in the literature for $m_{\rm DM}  \geq 150$ MeV for both DM annihilating and decaying in $e^+e^-$. For the $\mu^+\mu^-$ annihilation channel, we have the most stringent constraints for $m_{\rm DM} \geq 300$ MeV. For the remaining channels -- decays in $\mu^+\mu^-$, annihilations and decays in $\pi^+\pi^-$ -- we have the strongest bounds across the entire light DM mass range.

\medskip

In conclusion, including ICS emission is a powerful tool to compute strong bounds on light DM while circumventing the MeV gap left by old and current $\gamma$-ray observatories. Among the possibilities to improve our constraints, we could model the astrophysical background however at the cost of making them prone to additional uncertainties. Finally, future $\gamma$-ray observatories -- such as {\sc eAstrogam}, {\sc Amego} or {\sc Cosi} -- are expected to plug the MeV gap and therefore will be able to probe prompt emissions from annihilating or decaying light DM.


\newpage

\subsection*{Acknowledgments}
\bigskip

I am grateful to Marco Cirelli, Nicolao Fornengo, Elena Pinetti and Brandon M. Roach for their contribution to this work. I'm particularly thankful to Marco Cirelli for his comments on the proceeding. I also thank the organizers of the 34\textsuperscript{th} Rencontres de Blois in Particle Physics and Cosmology (Blois 2023), the XVIII International Conference on Topics in Astroparticle and Underground Physics (TAUP 2023) and TeV Particle Astrophysics (TeVPA) 2023 for giving me the opportunity to present this work.

\bibliographystyle{JHEP-noquotes}
\bibliography{XrayDM_proc.bib}

\end{document}